# Quantifying Long-Term Scientific Impact

Dashun Wang,[1,2]† Chaoming Song,[1,3]† and Albert-László Barabási[1,4,5,6]*

[1]Center for Complex Network Research, Department of Physics, Biology and Computer Science, Northeastern University, Boston, Massachusetts 02115, USA.

[2]IBM Thomas J. Watson Research Center, Yorktown Heights, New York 10598, USA.

[3]Department of Physics, University of Miami, Coral Gables, Florida 33124, USA

[4]Center for Cancer Systems Biology, Dana Farber Cancer Institute, Boston, Massachusetts 02115, USA

[5]Department of Medicine, Brigham and Women's Hospital, Harvard Medical School, Boston, Massachusetts 02115, USA.

[6]Center for Network Science, Central European University, Budapest, Hungary.

†These authors contributed equally to the work.

*Corresponding author. E-mail: alb@neu.edu

**Abstract**:

The lack of predictability of citation-based measures frequently used to gauge impact, from impact factors to short-term citations, raises a fundamental question: is there long-term predictability in citation patterns? Here we derive a mechanistic model for the citation dynamics of individual papers, allowing us to collapse the citation histories of papers from different journals and disciplines into a single curve, indicating that all papers tend to follow the same universal temporal pattern. The observed patterns not only help us uncover basic mechanisms that govern scientific impact, but also offer reliable measures of influence that may have potential policy implications.

Of the many tangible measures of scientific impact one stands out in its frequency of use: citations (*1–10*). The reliance on citation based measures, from the Hirsch index (*4*) to the g-index (*11*), from impact factors (*1*) to eigenfactors (*12*), and on diverse ranking based metrics (*13*), lies in the (often debated) perception that citations offer a quantitative proxy of a discovery's importance or a scientist's standing in the research community. Often lost in this debate is the fact that our ability to foresee lasting impact based on citation patterns has well-known limitations:

(i) The impact factor (IF) (*1*), conferring a journal's historical impact to a paper, is a poor predictor of a particular paper's future citations (*14, 15*): papers published in the same journal a decade later acquire widely different number of citations, from one to thousands (Fig. S2A).

(ii) The number of citations (*2*) collected by a paper strongly depends on the paper's age, hence citation-based comparisons favor older papers and established investigators. It also lacks predictive power: a group of papers that within a five year span collect the same number of citations are found to have widely different long-term impact (Fig. S2B).

(iii) Paradigm-changing discoveries have notoriously limited early impact (*3*), precisely because the more a discovery deviates from the current paradigm, the longer it takes to be appreciated by the community (*16*). Indeed, while for most papers their early and long-term citations correlate, this correlation breaks down for discoveries with the most long-term citations (Fig. 1B). Hence, publications with exceptional long-term impact appear to be the hardest to recognize on the basis of their early citation patterns.

(iv) Comparison of different papers is confounded by incompatible publication/citation/acknowledgement traditions of different disciplines and journals.

Long-term cumulative measures like the Hirsch index have predictable components, that can be extracted via data mining (*4, 17*). Yet, given the myriad of factors involved in the recognition of a new discovery, from the work's intrinsic value to timing, chance and the publishing venue, finding regularities in the citation history of individual papers, the minimal carriers of a scientific discovery, remains an elusive task.

In the past, much attention has focused on citation distributions, with debates on whether they follow a power law (*2, 18, 19*) or a log-normal form (*3, 7, 15*). Also, universality across disciplines allowed the rescaling of the distributions by discipline dependent variables (*7, 15*). Together, these results offer convincing evidence that the aggregated citation patterns are characterized by generic scaling laws. Yet, little is known about the mechanisms governing the temporal evolution of individual papers. The inherent difficulty in addressing this problem is well illustrated by the citation history of papers extracted from the Physical Review corpus (Fig. 1A), consisting of 463,348 papers published between 1893 and 2010 and spanning all areas of physics (*3*). The fat tailed nature of the citation distribution 30 years after publication indicates that while most papers are hardly cited, a few do have exceptional impact (Fig. 1B inset) (*2, 3, 7, 19, 20*). This impact heterogeneity, coupled with widely different citation histories (Fig. 1A), suggests a lack of order and hence lack of predictability in citation patterns. Yet, as we show next, this lack of order in citation histories is only apparent, as citations follow widely reproducible dynamical patterns that span research fields.

We start by identifying three fundamental mechanisms that drive the citation history of individual papers:

A) *Preferential attachment* captures the well-documented fact that highly cited papers are more visible and are more likely to be cited again than less-cited contributions (*20, 21*). Accordingly a paper *i*'s probability to be cited again is proportional to the total number of citations $c_i$ the paper received previously (Fig. S3).

B) *Aging* captures the fact that new ideas are integrated in subsequent work, hence each paper's novelty fades eventually (*22, 23*). The resulting long term decay is best described by a log-normal survival probability (see Fig. 1C and SOM S2.1)

$$P_i(t) = \frac{1}{\sqrt{2\pi}\sigma_i t} \exp\left(-\frac{(\ln t - \mu_i)^2}{2\sigma_i^2}\right) \qquad (1)$$

C) *Fitness*, $\eta_i$, captures the inherent differences between papers, accounting for the perceived novelty and importance of a discovery (*24, 25*). Novelty and importance depend on so many intangible and subjective dimensions that it is impossible to objectively quantify them all. Here we bypass the need to evaluate a paper's intrinsic value and view fitness $\eta_i$ as a collective measure capturing the community's response to a work.

Combining A–C, we can write the probability that paper *i* is cited at time *t* after publication as

$$\Pi_i(t) \sim \eta_i c_i^t P_i(t). \qquad (2)$$

Solving the associated master equation, Eq. 2 allows us to predict the cumulative number of citations acquired by paper $i$ at time $t$ after publication (SOM S2.2)

$$c_i^t = m\left(e^{\frac{\beta\eta_i}{A}\Phi\left(\frac{\ln t - \mu_i}{\sigma_i}\right)} - 1\right) \equiv m\left(e^{\lambda_i\Phi\left(\frac{\ln t - \mu_i}{\sigma_i}\right)} - 1\right), \qquad (3)$$

where

$$\Phi(x) \equiv (2\pi)^{-1/2}\int_{-\infty}^{x} e^{-y^2/2}dy \qquad (4)$$

is the cumulative normal distribution, $m$ measures the average number of references each new paper contains, $\beta$ captures the growth rate of the total number of publications (SOM S1.3) and $A$ is a normalization constant (SOM S2.2). Hence $m$, $\beta$ and $A$ are global parameters, having the same value for all publications. We have chosen $m=30$ throughout the paper, as our results do not depend on this choice (SOM S2.3). Equation 3 represents a minimal citation model, that captures all known quantifiable mechanisms that affect citation histories. It predicts that the citation history of paper $i$ is characterized by three fundamental parameters: the *relative fitness* $\lambda_i \equiv \eta_i\beta/A$, capturing a paper's importance relative to other papers; the *immediacy* $\mu_i$, governing the time for a paper to reach its citation peak and the *longevity* $\sigma_i$, capturing the decay rate. Using the rescaled variables $\tilde{t} \equiv (\ln t - \mu_i)/\sigma_i$ and $\tilde{c} \equiv \ln(1 + c_i^t/m)/\lambda_i$, we obtain our main result,

$$\tilde{c} = \Phi(\tilde{t}), \qquad (5)$$

predicting that each paper's citation history should follow the same universal curve $\Phi(\tilde{t})$ if rescaled with the paper-specific $(\lambda_i, \mu_i, \sigma_i)$ parameters. Therefore, given a paper's citation history, i.e. $t$ and $c_i^t$, we can obtain the best-fitted three parameters for paper $i$ using Eq. 3. To illustrate the process, we selected a paper from our corpus, whose citation history is shown in Fig. 1D,E. We fit to Eq. 3 the paper's cumulative citations (Fig. 1E) using the least square fit method, obtaining $\lambda = 2.87$, $\mu = 7.38$ and $\sigma = 1.2$. To illustrate the validity of the fit, in Fig. 1E we show the prediction of Eq. 3 using the uncovered fit parameters.

To test the model's validity, we rescaled all papers published between 1950 and 1980 in the Physical Review corpus, finding that they all collapse into Eq. 5 (Fig. 1F, see also SOM S2.4.1 for the statistical test of the data collapse). The reason is explained in Fig. 1G: by varying λ, μ and σ, Eq. 3 can account for a wide range of empirically observed citation histories, from jump-decay patterns to delayed impact. We also tested our model on all papers published in 1990 by 12 prominent journals (Table S4), finding an excellent collapse for all (see Fig. 1G inset for Science and SOM S2.4.2 and Fig. S8 for the other journals).

The model Eqs. 3-5 also predicts several fundamental measures of impact:

*Ultimate impact* ($c^\infty$) represents the total number of citations a paper acquires during its lifetime. By taking the $t \to \infty$ limit in Eq. 3, we obtain

$$c_i^\infty = m(e^{\lambda_i} - 1), \tag{6}$$

a simple formula that predicts that the total number of citations acquired by a paper during its lifetime is independent of immediacy ($\mu$) or the rate of decay ($\sigma$), and *depends only on a single parameter, the paper's relative fitness*, $\lambda$.

*Impact time* ($T_i^*$) represents the characteristic time it takes for a paper to collect the bulk of its citations. A natural measure is the time necessary for a paper to reach the geometric mean of its final citations, obtaining (SOM S2.2)

$$T_i^* \approx \exp(\mu_i). \tag{7}$$

Hence impact time is mainly determined by the immediacy parameter $\mu_i$ and is independent of fitness $\lambda_i$ or decay $\sigma_i$.

The proposed model offers a journal free methodology to evaluate long term impact. To illustrate this we selected three journals with widely different IFs: *Physical Review B* (*PRB*) (IF = 3.26 in 1992), *PNAS* (10.48) and *Cell* (33.62), and measured for each paper published by them the fitness $\lambda$, obtaining their distinct journal-specific $P(\lambda)$ fitness distribution (Fig. 2A). We then selected all papers with comparable fitness $\lambda \approx 1$, and followed their citation histories. As expected they follow different paths: *Cell* papers ran slightly ahead and *PRB* papers stay behind, resulting in distinct $P(c^T)$ distributions for years $T = 2 \div 4$. Yet, by year 20 the cumulative number of citations acquired by these papers shows a remarkable convergence to each other (Fig. 2B), supporting our prediction that given their similar fitness $\lambda$, eventually they will have the same ultimate impact $c^\infty$=51.5. To quantify the magnitude of the observed convergence, we measured the coefficient of variation $\sigma_c/\langle c \rangle$ for $P(c^T)$, finding that this ratio decreases with time (Fig. 2C). This helps us move beyond visual inspection, offering quantitative evidence that in the long run the differences in citation counts between these papers vanishes with time, as predicted by our model. In contrast, if we choose all papers with the same number of citations at year two (i.e. the same $c^2$, Fig. 2D), the citations acquired by them diverge with time and $\sigma_c/\langle c \rangle$ increases (Fig. 2E,F), supporting our conclusion that these quantities lack predictability. Therefore $\lambda$ and $c^\infty$ offer a journal independent measure of a publication's long-term impact.

The model (Eqs. 3–5) also helps connect the impact factor, the traditional measure of impact of a scientific journal, to the journal's $\Lambda$, $M$, and $\Sigma$ parameters (the analogs of $\lambda$, $\mu$, $\sigma$, S4),

$$IF \approx \frac{m}{2}\left(\exp\left[\Lambda\Phi\left(\frac{M_1-M}{\Sigma}\right)\right] - \exp\left[\Lambda\Phi\left(\frac{M_2-M}{\Sigma}\right)\right]\right). \tag{8}$$

Knowing $\Lambda$, in analog with (6) we can calculate a journal's ultimate impact as $C^\infty = m(e^\Lambda - 1)$, representing the total number of citations a paper in the journal will receive during its lifetime. As we show in the SOM S4, Eq. 8 predicts a journal's impact factor in good agreement with the values reported by ISI. Equally important, it helps us understand the mechanisms that influence the evolution of the IF, as illustrated by the changes in the impact factor of *Cell* and *NEJM*. In 1998 the IFs of *Cell* and *NEJM* were 38.7 and 28.7, respectively (Fig. 3A). Yet over the next decade there was a remarkable reversal: *NEJM* became the first journal to reach IF = 50, while *Cell*'s IF decreased to around 30. This raises a puzzling question: has the impact of papers published by the two journals changed so dramatically? To answer this we determined $\Lambda$, M, and $\Sigma$ for both journals from 1996 to 2006 (Fig. 3D–F). While $\Sigma$ were indistinguishable (Fig. 3D), we find that the fitness of NEJM increased from $\Lambda = 2.4$ (1996) to $\Lambda = 3.33$ (2005), increasing the journal's ultimate impact from $C^\infty = 300$ (1996) to a remarkable $C^\infty = 812$ (2005) (Fig. 3B). But *Cell*'s $\Lambda$ also increased in this period (Fig. 3E), moving its ultimate impact from $C^\infty = 366$ (1996) to 573 (2005). Yet, if both journals attracted papers with increasing long-term impact, why did *Cell*'s IF drop and *NEJM*'s grow? The answer lies in changes in the impact time $T^*=\exp(M)$: while *NEJM*'s impact time remained unchanged at $T^* \approx 3$ years, *Cell*'s $T^*$ increased from $T^* = 2.4$ years to $T^* = 4$ years (Fig. 3C). Therefore, *Cell* papers have gravitated from short to long-term impact: a typical *Cell* paper gets 50% more citations than a decade ago, but fewer of the citations come within the first two years (Fig. 3C, inset). In contrast, with a largely unchanged $T^*$, *NEJM*'s increase in $\Lambda$ translated into a higher IF. These conclusions are fully supported by the $P(\lambda)$ and $P(\mu)$ distributions for individual papers published by *Cell* and *NEJM* in 1996 and 2005: both journals show a clear shift to higher fitness papers (Fig. 3G), but while $P(\mu)$ is largely unchanged for *NEJM*, there is a clear shift to higher $\mu$ papers in *Cell* (Fig. 3H).

Can we use the developed framework to predict the future citations of a publication? For this we adopt a framework borrowed from weather predictions and data mining: we use paper $i$'s citation history up to year $T_{Train}$ after publication (training period) to estimate $\lambda_i$, $\mu_i$, $\sigma_i$ and then use the model Eq. 3 to predict its future citations $c_i^t$ and Eq. 6 to determine its ultimate impact $c_i^\infty$. Yet, the uncertainties in estimating $\lambda_i$, $\mu_i$, $\sigma_i$ from the inherently noisy citation histories affect our predictive accuracy (see SOM S2.6). Hence instead of simply interpolating Eq. 3 into the future, we assign a citation envelope to each paper, explicitly quantifying the uncertainty of our predictions (see S2.6). In Fig. 4A, we show the predicted most likely citation path (red line) with the uncertainty envelope (grey area) for three papers, based on a 5 year training period. Two of the three papers fall within the envelope, for the third, however, the model overestimated the future citations. Increasing the training period enhanced the predictive accuracy (Fig. 4B).

To quantify the model's overall predictive accuracy we measured the fraction of papers that fall within the envelope for all PR papers published in 1960s. That is, we measured the $z_{30}$-score for each paper, capturing the number of standard deviations $z_{30}$ the real citations $c^{30}$ deviate from the most likely citation 30 years after publication. The obtained $P(z_{30})$ distribution across all papers decayed fast with $z_{30}$ (Fig. 4C), indicating that large $z$ values are extremely rare. With $T_{Train} = 5$ only 6.5% of the papers left the prediction envelope 30 years later, hence the model correctly approximated the citation range for 93.5% of papers 25 years into the future.

The observed accuracy prompts us to ask whether the proposed model is unique in its ability to capture future citation histories. We therefore identified several models that have been either used in the past to fit citation histories, or have the potential to do so: the Logistic (*26*), Bass (*27*), and Gompertz (*26*, *28*) models (for formulae see SOM, Table S2)

We fit the predictions of these models to PR papers and used the weighted Kolmogorov-Smirnov (KS) test to evaluate their goodness of fit (see Eq. S43 for definition), capturing the maximum deviation between the fitted and the empirical data. The lowest KS distribution across most papers was observed with Eq. 3, indicative of the best fit (Fig. 4D). The reason is illustrated in Fig. S18: the symmetric $c(t)$ predicted by the Logistic Model cannot capture the asymmetric citation curves. While the Gompertz and the Bass models predict asymmetric citation patterns, they also predict an exponential (Bass) or double-exponential (Gompertz) decay of citations (Table S2), much faster than observed in real data. To quantify how these deviations affect the predictive power of each of these models, we used a 5 and a 10 year training period to fit the parameters of each model and computed the predicted most likely citations at year 30 (Fig. 4E,F). Independent of the training period the predictions of the Logistic, Bass and Gompertz models always lay outside the 25%–75% prediction quartiles (red bars), systematically underestimating future citations. In contrast, the prediction of Eq. 3 for both training periods was within the 25-75% quantiles, its accuracy visibly improving for the ten year training period (Fig. 4F). In Supplementary Materials S3.3 we offer additional quantitative assessment of these predictions (Fig. S19), demonstrating our model's predictive power pertaining to both the fraction of papers whose citations it correctly predicts and in the magnitude of deviations between the predicted and the real citations. The predictive limitations of the current models was also captured by their $P(z_{30})$ distribution, indicating that for the Logistic, Bass and Gompertz model more than half of the papers underestimate with more than two standard deviations the true citations ($z > 2$) at year 30 (Fig. 4C), in contrast with 6.5% for the proposed model (Eq. 3).

Ignoring preferential attachment in Eq. 2 leads to the *Lognormal model*, containing a lognormal temporal decay modulated by a single fitness parameter. As we analytically show in S3.4, for small fitness Eq. 3 converged to the Lognormal model, which correctly captured the citation history of small impact papers. The Lognormal model failed, however, to predict the citation patterns of medium to high impact papers (Fig. S20). The proposed model therefore allows us to analytically predict the citation threshold when preferential attachment becomes relevant. The calculations indicate that the Lognormal model is indistinguishable from the predictions of Eq. 3 for papers that satisfy the equation

$$\sum_{n=2}^{\infty} \frac{1}{n!} \Phi^n \lambda^n < 1. \tag{9}$$

Solving this equation predicts $\lambda < 0.25$, equivalent with the citation threshold $c^\infty < 8.5$, representing the theoretical bound for preferential attachment to turn on. This analytical prediction is in excellent agreement with empirical finding that preferential attachment is masked by initial attractiveness for papers with less than seven citations (*29*). Note that the lognormal function has been proposed before to capture the citation distribution of a body of papers (*15*). Yet, the lognormals appearing in Ref (*15*) and in the Lognormal model discussed above have different origins and implications (SOM S2.5.2).

The proposed model has obvious limitations: it cannot account for exogenous "second acts", like the citation bump observed for superconductivity papers following the discovery of high temperature superconductivity in the 1980s, or delayed impact, like the explosion of citations to Erdős and Rényi's work four decades after their publication, following the emergence of network science (*3, 20, 21, 23*).

Our findings have policy implications, as current measures of citation-based impact, from IF to Hirsch index (*4, 17*), are frequently integrated in reward procedures, the assignment of research grants, awards and even salaries and bonuses (*30*), despite their well-known lack of predictive power. In contrast with the IF and short-term citations that lack predictive power, we find that $c^\infty$ offers a journal-independent assessment of a paper's long term impact, with a meaningful interpretation: it captures the total number of citations a paper will ever acquire, or the discovery's ultimate impact. While additional variables combined with data mining could further enhance the demonstrated predictive power, an ultimate understanding of long-term impact will benefit from a mechanistic understanding of the factors that govern the research community's response to a discovery.

# References and Notes

1. E. Garfield. The history and meaning of the journal impact factor. *JAMA: the journal of the American Medical Association*, 295(1):90–93, 2006.
2. D.J. de Solla Price. Networks of scientific papers. *Science*, 149(3683):510–515, 1965.
3. S. Redner. Citation statistics from 110 years of physical review. *Physics Today*, 58:49, 2005.
4. J.E. Hirsch. An index to quantify an individual's scientific research output. *Proceedings of the National Academy of Sciences of the United states of America*, 102(46):16569, 2005.
5. S. Lehmann, A.D. Jackson, and B.E. Lautrup. Measures for measures. *Nature*, 444(7122): 1003–1004, 2006.
6. B.F. Jones, S. Wuchty, and B. Uzzi. Multi-university research teams: shifting impact, geography, and stratification in science. *Science*, 322(5905):1259–1262, 2008.
7. F. Radicchi, S. Fortunato, and C. Castellano. Universality of citation distributions: Toward an objective measure of scientific impact. *Proceedings of the National Academy of Sciences*, 105(45):17268–17272, 2008.
8. J.A. Evans and J. Reimer. Open access and global participation in science. *Science*, 323 (5917):1025, 2009.
9. J.A. Evans and J.G. Foster. Metaknowledge. *Science*, 331(6018):721–725, 2011.
10. A.-L. Barabási, C. Song, and D. Wang. Publishing: Handful of papers dominates citation. Nature, 491(7422):40, 2012.
11. L. Egghe. Theory and practise of the g-index. *Scientometrics*, 69(1):131–152, 2006.
12. A. Fersht. The most influential journals: Impact factor and eigenfactor. *Proceedings of the National Academy of Sciences*, 106(17):6883–6884, 2009.
13. F. Radicchi, S. Fortunato, B. Markines, and A. Vespignani. Diffusion of scientific credits and the ranking of scientists. *Physical Review E*, 80(5):056103, 2009.
14. P.O. Seglen. Why the impact factor of journals should not be used for evaluating research. *BMJ: British Medical Journal*, 314(7079):498, 1997.
15. M.J. Stringer, M. Sales-Pardo, and L.A.N. Amaral. Effectiveness of journal ranking schemes as a tool for locating information. *PLoS ONE*, 3(2):e1683, 02 2008.
16. T.S. Kuhn. *The structure of scientific revolutions*. University of Chicago press, 1996.
17. D.E. Acuna, S. Allesina, and K.P. Kording. Future impact: Predicting scientific success. *Nature*, 489(7415):201–202, 2012.
18. G.J. Peterson, S. Pressé, and K.A. Dill. Nonuniversal power law scaling in the probability distribution of scientific citations. *Proceedings of the National Academy of Sciences*, 107(37): 16023–16027, 2010.
19. S. Redner. How popular is your paper? an empirical study of the citation distribution. *The European Physical Journal B*, 4(2):131–134, 1998.
20. A.-L. Barabási and R. Albert. Emergence of scaling in random networks. *Science*, 286(5439): 509–512, 1999.

21. G. Caldarelli. *Scale-Free Networks*. Oxford University Press, 2007.
22. M. Medo, G. Cimini, and S. Gualdi. Temporal effects in the growth of networks. *Physical Review Letters*, 107(23):238701, 2011.
23. S.N. Dorogovtsev and J.F.F. Mendes. *Evolution of networks: From biological nets to the Internet and WWW*. Oxford, 2003.
24. G. Bianconi and A.-L. Barabási. Competition and multiscaling in evolving networks. *EPL (Europhysics Letters)*, 54:436, 2001.
25. G. Caldarelli, A. Capocci, P. De Los Rios, and M.A. Muñoz. Scale-free networks from varying vertex intrinsic fitness. *Physical Review Letters*, 89(25):258702, 2002.
26. V. Mahajan, E. Muller, and F.M. Bass. New product diffusion models in marketing: A review and directions for research. *The Journal of Marketing*, pages 1–26, 1990.
27. F.M. Bass. Comments on "a new product growth for model consumer durables the bass mode". *Management science*, 50(12):1833–1840, 2004.
28. B. Gompertz. On the nature of the function expressive of the law of human mortality, and on a new mode of determining the value of life contingencies. *Philosophical transactions of the Royal Society of London*, 115:513–583, 1825.
29. Y.H. Eom and S. Fortunato. Characterizing and modeling citation dynamics. *PloS one*, 6(9): e24926, 2011.
30. I. Fuyuno and D. Cyranoski. Cash for papers: Putting a premium on publication. *Nature*, 441(7095):792, 2006.

**Acknowledgements**: The authors wish to thank P. Azoulay, C. Hidalgo, J. Loscalzo, D. Pedreschi, B. Uzzi, M. Vidal, and members of CCNR for insightful discussions. We wish to thank the anonymous Referee for suggesting the Lognormal model, which led to the analytical prediction of the citation threshold for preferential attachment. The PR dataset is available upon request through American Physical Society. Supported by Lockheed Martin Corporation SRA 11.18.11, NSCTA sponsored by the US Army Research Laboratory under Agreement Number W911NF-09-2-0053, DARPA under Agreement No. 11645021, and the Future and Emerging Technologies Project Nr 317 532 "Multiplex" financed by the European Commission.

Figure 1: **Characterizing citation dynamics**. (A) Yearly citation $c_i(t)$ for 200 randomly selected papers published between 1960 and 1970 in the Physical Review (PR) corpus. The color code corresponds to each papers' publication year. (B) Average number of citations acquired two years after publication ($c^2$) for papers with the same long-term impact ($c^{30}$), indicating that for high impact papers ($c^{30} \geq 400$, shaded area) the early citations underestimate future impact. Inset: Distribution of citations 30 years after publication ($c^{30}$) for PR papers published between 1950 and 1980. (C) Distribution of papers' age when they get cited. To separate the effect of preferential attachment, we measure the aging function for papers with the same number of previous citations (here $c^t = 20$, see also S2.1). The solid line corresponds to Gaussian fit of the data, indicating $P(\ln \Delta t | c^t)$ follows a normal distribution. (D) Yearly citation $c(t)$ for a research paper from the PR corpus. (E) Cumulative citations $c^t$ for the paper in (D) together with the best fit to Eq. 3 (solid line). (F) Data collapse for 7,775 papers with more than 30 citations within 30 years in the PR corpus published between 1950 and 1980. Inset: data collapse for the 20 years citation histories of all papers published by Science in 1990 (842 papers). (G) Changes in the citation history $c(t)$ according to Eq. 3 after varying the $\lambda, \mu, \sigma$) parameters, indicating that Eq. 3 can account for a wide range of citation patterns.

Figure 2: **Evaluating long-term Impact**. (A) Fitness distribution $P(\lambda)$ for papers published by *Cell*, *PNAS*, and *Physical Review B* (*PRB*) in 1990. Shaded area indicates papers in the $\lambda \approx 1$ range selected for further study. (B) Citation distributions for papers with fitness $\lambda \approx 1$ highlighted in (A) for years 2, 4, 10, and 20 after publication. (C) Time dependent relative variance of citations for papers selected in (A). (D) Citation distribution two years after publication ($P(c^2)$) for papers published by *Cell*, *PNAS*, and *PRB*. Shaded area highlights papers with $c^2 \in [5,9]$ selected for further study. (E) Citation distributions for papers with $c^2 \in [5,9]$ selected in (D) after 2, 4, 10, and 20 years. (F) Time dependent relative variance of citations for papers selected in (D).

Figure 3: **Quantifying changes in a journal's long-term impact**. (A) Impact factor of *Cell* and *New England Journal of Medicine* (*NEJM*) reported by Thomson Reuters from 1998 to 2006. (B) Ultimate impact $C^\infty$ (see Eq. 6) of papers published by the two journals from 1996 to 2005. (C) Impact time $T^*$ (Eq. 7) of papers published by the two journals from 1996 to 2005. Inset: fraction of citations that contribute to the IF. (D–F)

The measured time dependent longevity (Σ), fitness (Λ), and immediacy (M) for the two journals. (G) Fitness distribution for individual papers published by *Cell* (left) and *NEJM* (right) in 1996 (black) and 2005 (red). (H) Immediacy distributions for individual papers published by *Cell* (left) and *NEJM* (right) in 1996 (black) and 2005 (red).

Figure 4: **Predicting Future Citations**. (A, B) Prediction envelope for three papers obtained using a five (A) and ten (B) years of training (shaded vertical area). The middle curve offers an example of a paper for which the prediction envelope misses the future evolution of the citations. The envelope illustrates the range for which $z \leq 1$. Comparing A and B illustrates how the increasing training period decreases the uncertainty of the prediction, resulting in a narrower envelope. (C) Complementary cumulative distribution of $z_{30}$ ($P^{>}(z_{30})$), where $z_{30}$ quantifies how many standard deviations the predicted citation history deviates from the real citation curve thirty years after publication (see also S2.6). We selected papers published in 1960s in the PR corpus that acquire at least 10 citations in 5 years (4,492 in total). The red curve captures predictions for 30 years after publication for $T_{Train} = 10$, indicating that for our model 93.5% papers have $z_{30} \leq 2$. The blue curve relies on 5 year training. The grey curves capture the predictions of Gompertz, Bass, and Logistic model for 30 years after publication by using 10 years as training. (D) Goodness of fit using weighted Kolmogorov-Smirnov (KS) test (see S3.3), indicating that Eq. (3) offers the best fit to our testing base (same as the papers in C) (E, F) Scatter plots of predicted citations and real citations at year 30 for our test base (same sample as in C, D), using as training data the citation history for the first 5 (E) or 10 (F) years. The error bars indicate prediction quartiles (25% and 75%) in each bin, and are colored green if $y = x$ lies between the two quartiles in that bin, and red otherwise. The black circles correspond to the average predicted citations in that bin.

**Supplementary Materials**

**Materials and Methods**

**Supplementary Text**

**Tables S1 to S4**

**Figs. S1 to S25**

**References (*31–44*)**

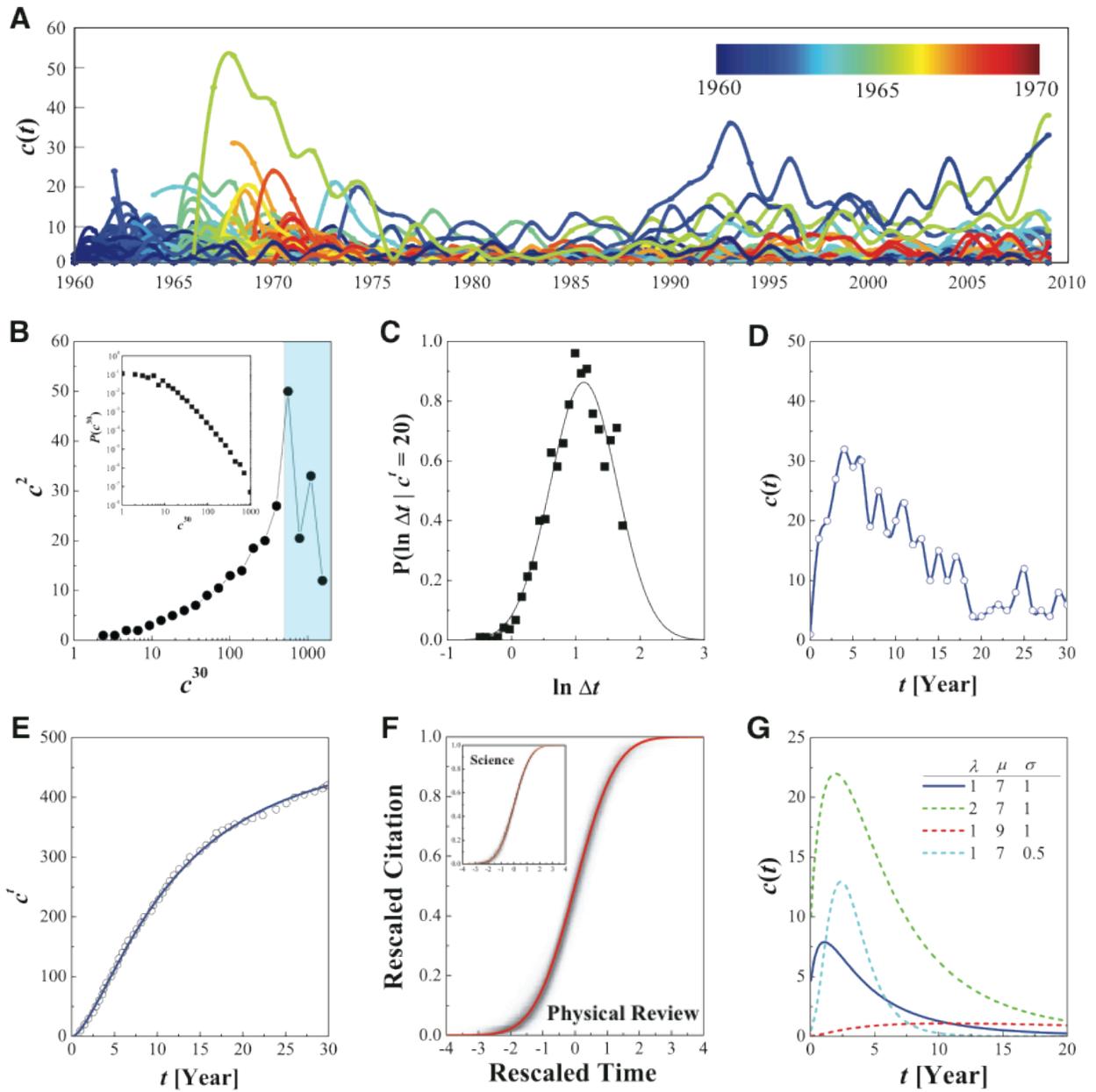

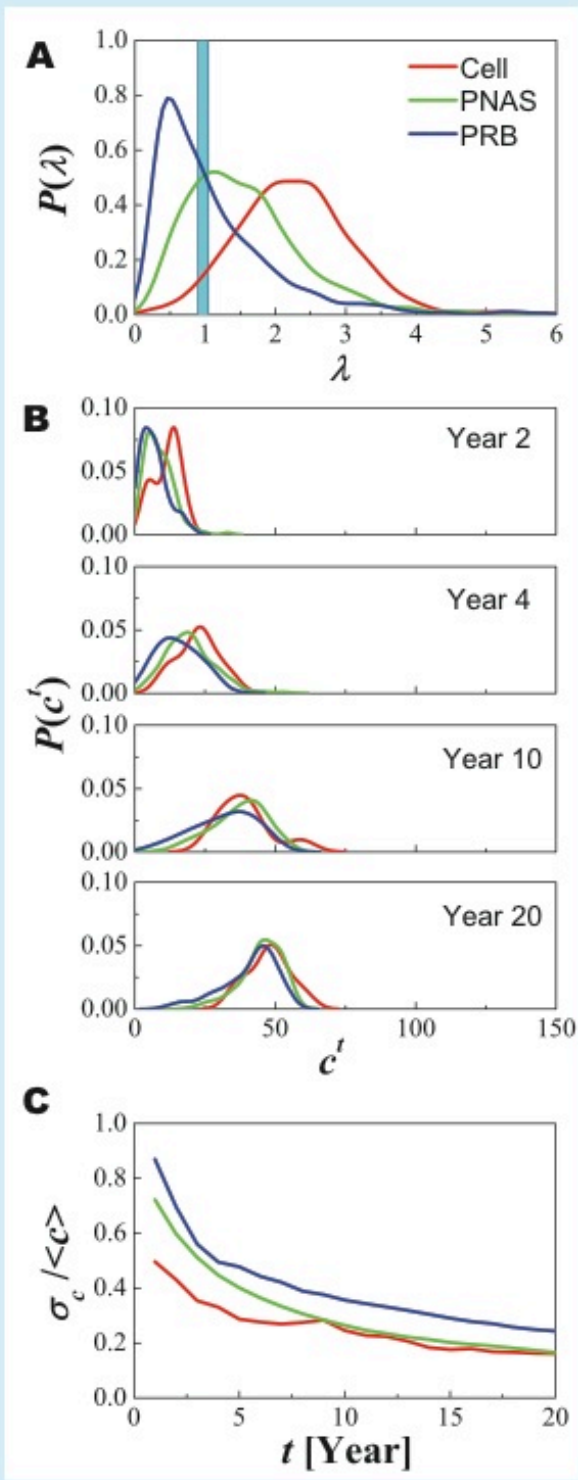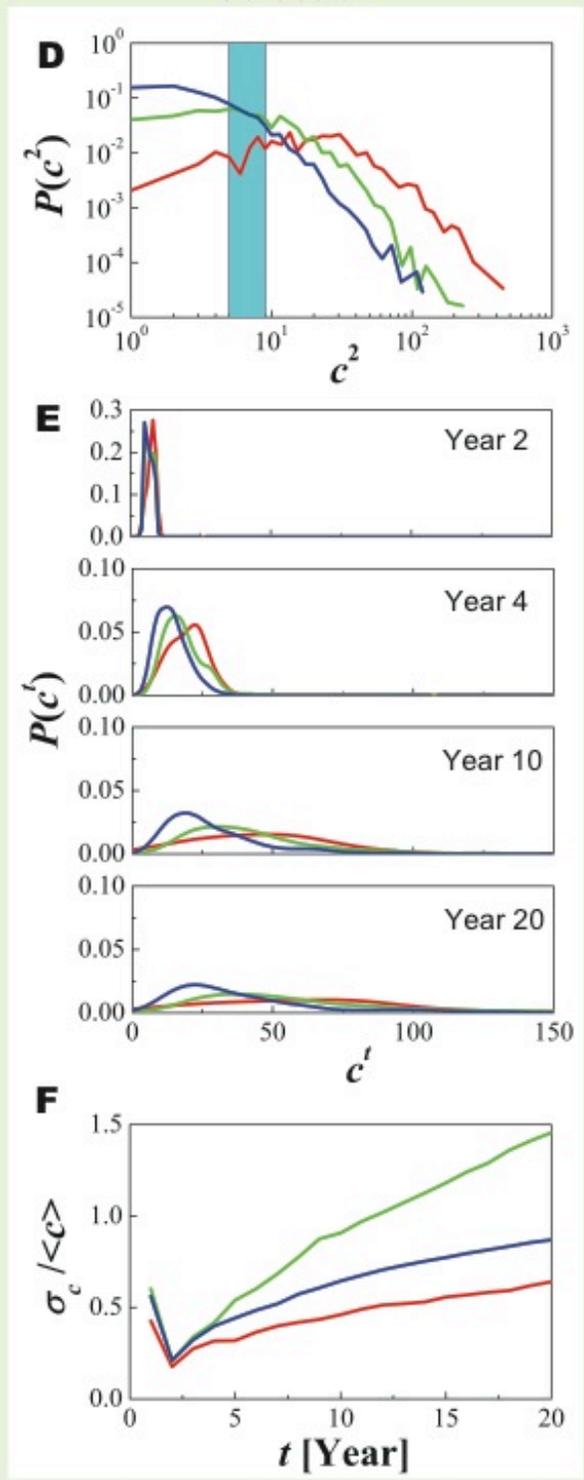

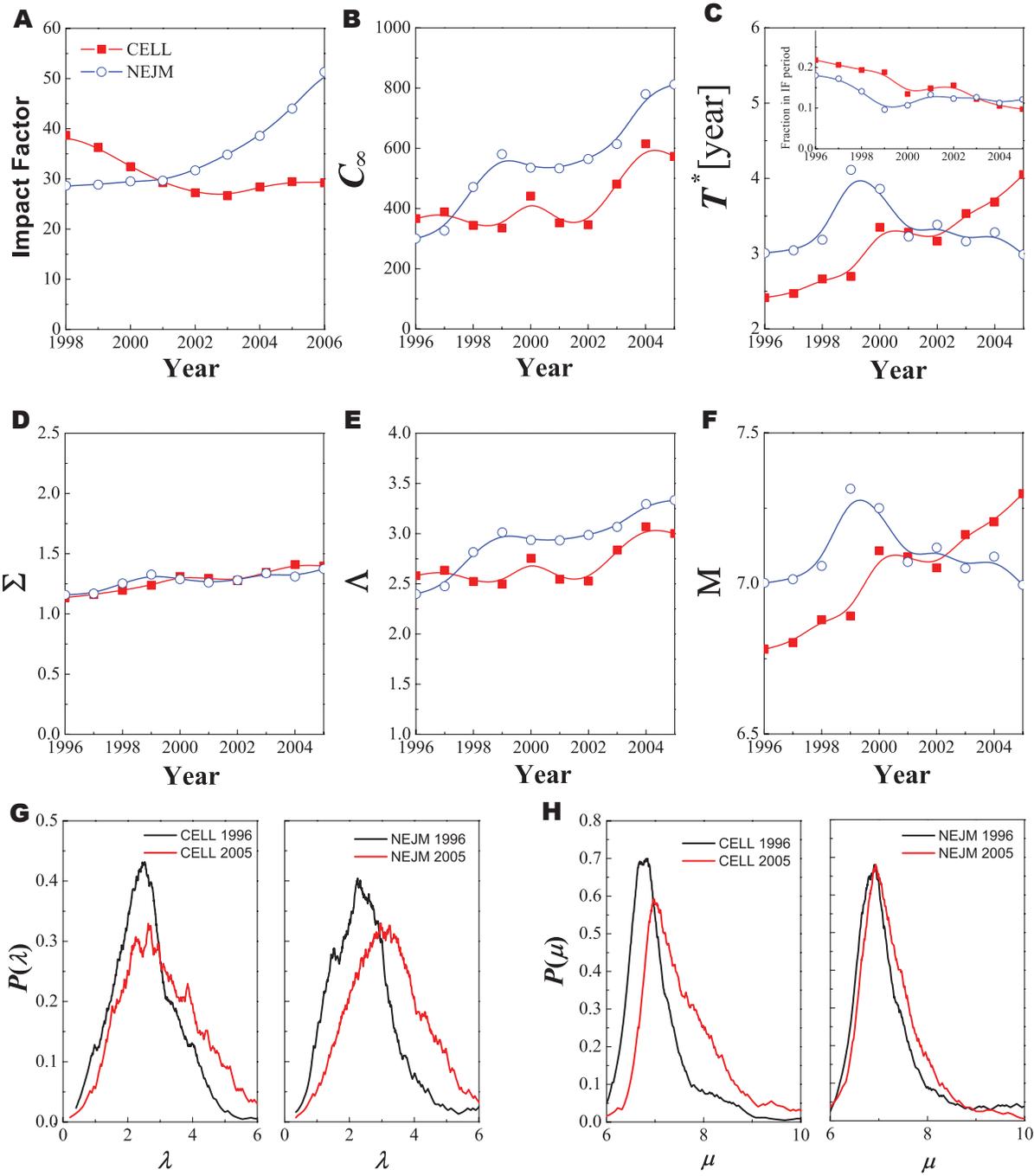

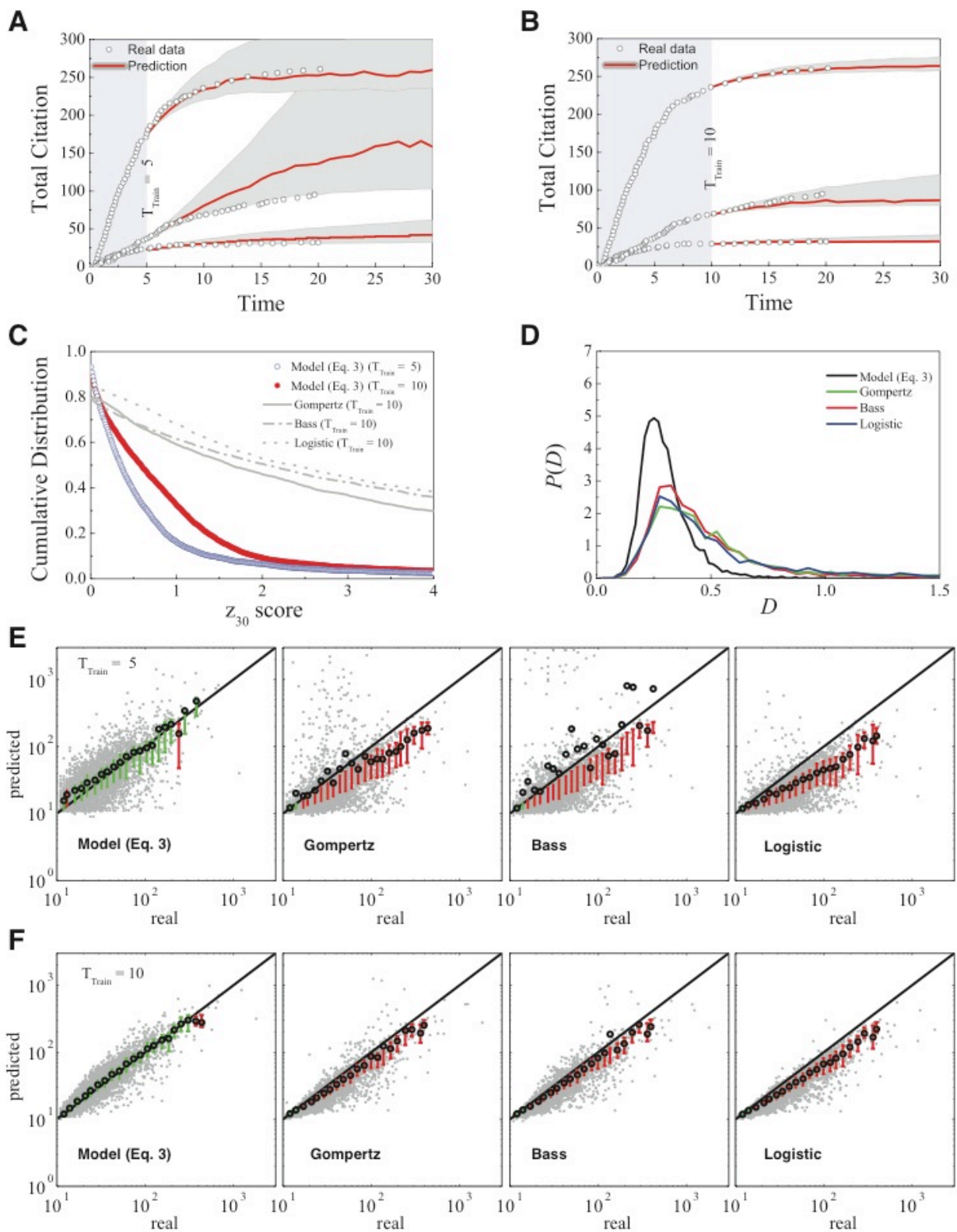